This figure "m87fig3.jpg" is available in "jpg" format from:

http://arxiv.org/ps/astro-ph/0204103v1

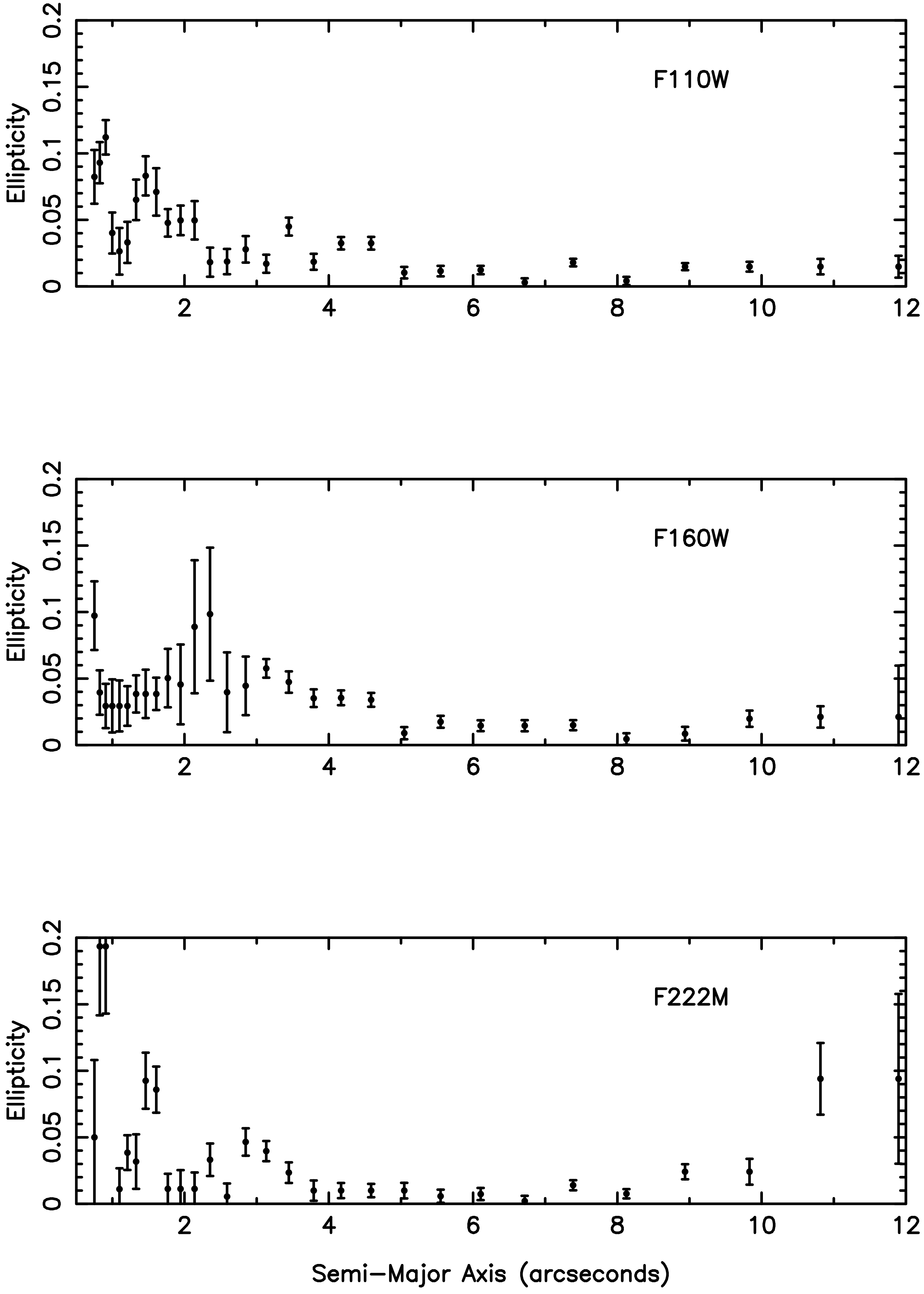

This figure "m87fig5.jpg" is available in "jpg" format from:

http://arxiv.org/ps/astro-ph/0204103v1

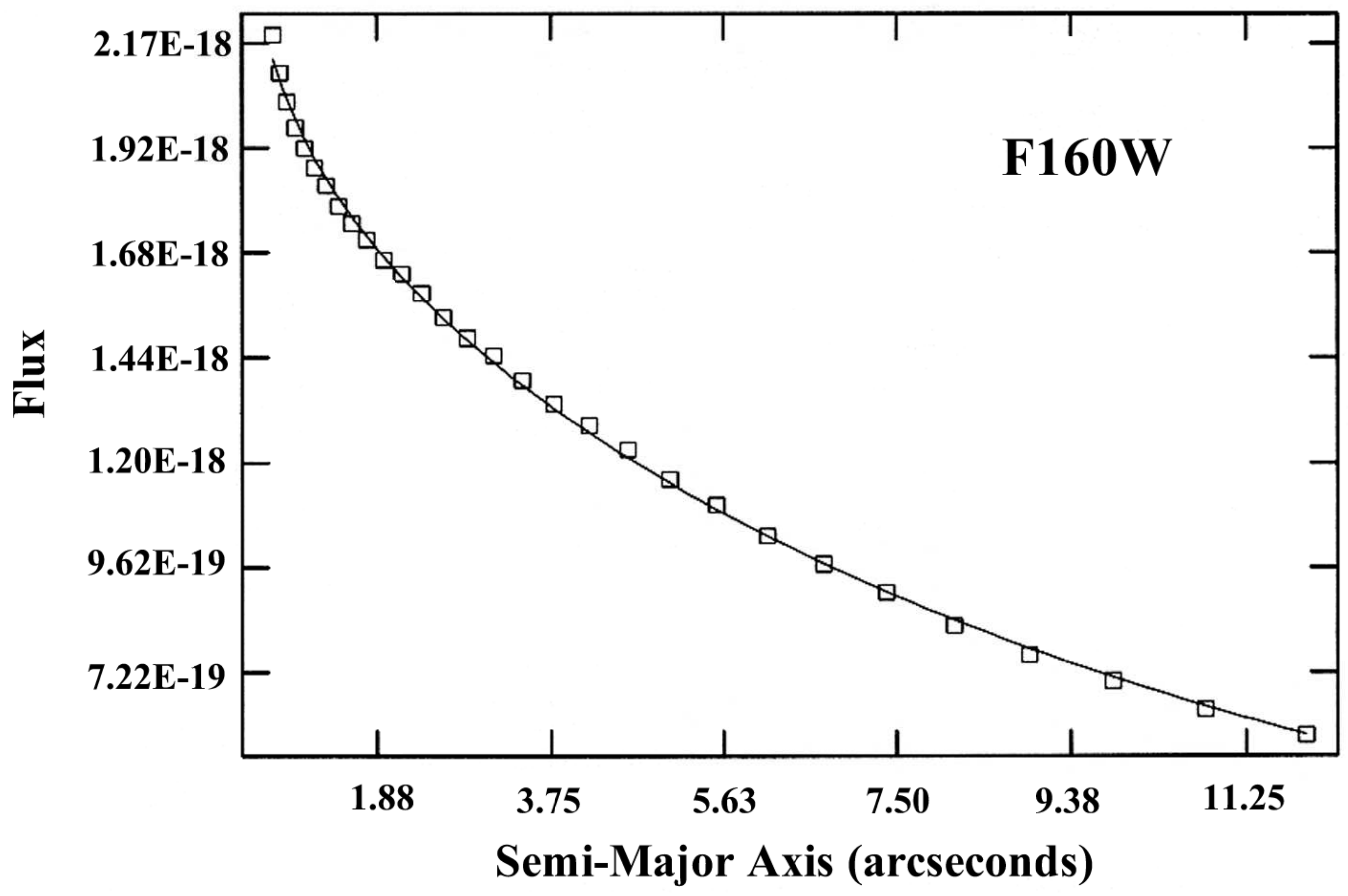

# *Hubble Space Telescope* NICMOS Imaging of the Core of M87

Michael R. Corbin[1], Earl O'Neil & Marcia J. Rieke

NICMOS Group, Steward Observatory, The University of Arizona,

Tucson, AZ 85721

[1]Current address : Computer Sciences Corporation, Space Telescope Science Institute,

3700 San Martin Drive, Baltimore, MD 21218; contact: corbin@stsci.edu

## ABSTRACT

We present broad-band 1.1, 1.6 and 2.2 μm images and a 2.37 μm narrow-band image of the inner 19″ of the nearby radio galaxy M87, obtained with the Near Infrared Camera and Multi-Object Spectrometer of the *Hubble Space Telescope* (HST). The isophotes of the broad-band images are almost perfectly circular to within approximately 0.5″ (~ 50 pc) of the active nucleus, and an $r^{1/4}$-law provides a good fit to the galaxy brightness profile at these wavelengths to within the same distance. This result agrees with predictions that the nuclear supermassive black hole will produce a nearly spherical distribution of the surrounding stars within a galaxy crossing time. A difference image formed from the 1.6 μm image and a *V* -band image obtained with the HST Wide Field Planetary Camera 2 does not show any clear evidence of a physically thick dusty torus around the nucleus, consistent with its lack of strong thermal mid-infrared emission. If such a torus is present, our data indicate its outer radius to be < 50 pc. The infrared broadband colors and 2.37 μm image (which is sensitive to the strength of the stellar CO absorption) show no gradients to a distance of approximately 5″ (~ 400 pc) from the nucleus, and are consistent with a population dominated by late M giants, with no evidence of recent star formation. However, the globular clusters in this region are confirmed to consist of stars bluer than the underlying galaxy, indicating a different formation history. The images and associated colors also confirm that the regions beyond the nucleus do not contain strongly concentrated dust, in contrast to many other radio galaxies. In combination with other recent observations, these results indicate that M87 represents the dynamically evolved product of past galaxy mergers, and suggest that its nucleus is in the final stages of activity.

## 1. INTRODUCTION

*Hubble Space Telescope* (HST) spectroscopy of M87 has provided strong evidence that its active nucleus is powered by accretion onto a supermassive (~ 2 - 4 × $10^9$ $M_\odot$) black hole (Harms et al. 1994; Marconi et al. 1997; Macchetto et al. 1997). This spectroscopic evidence is however based on the kinematics of a complex set of ionized and moderately dusty filaments surrounding the nucleus, and extending ~ 2′ away from it (Sparks, Ford & Kinney 1993; Ford et al. 1994; Sparks et al. 2000). The influence of the central black hole on the surrounding stars is consequently difficult to judge from optical data, although the presence of a cusp in the nuclear brightness profile has already been noted (Young et al. 1978; Lauer et al. 1992). More recently, Merritt & Quinlan (1998) have found that a ~ $10^9$ $M_\odot$ black hole should dominate the stellar kinematics of M87 to a distance ~ 250 pc from the nucleus, which corresponds to a scale of approximately 3″, easily accessible to HST[1]. Testing such models is best accomplished by observations in the near infrared, where the optical depth of dust is lower than at optical wavelengths. The near infrared region is also relatively free from strong nebular line emission, such as in [O III] λλ4959, 5007 and Hα. Similarly, recent kinematic mapping of the M87 core region has revealed irregularities that suggest that M87 formed via the merger(s) of smaller galaxies (Pierce & Berrington 2000), like some other giant ellipticals. If the nuclear black hole of M87 were formed from a coalescence of two or more black holes acquired in such mergers, the formation process might also produce "stellar wakes" with isophotal signatures that would also best be seen in the near infrared (see Faber et al. 1997). M87 also shows a relative deficiency of dust compared to other galaxies in its class (Sparks et al. 2000), which can be further investigated by measurement of its infrared colors.

---

[1] Following Cohen (2000) we adopt a distance to M87 of 16.3 Mpc, which is a weighted average of the estimates of Whitmore et al. (1995) and Pierce et al. (1994). The corresponding scale is 75 pc $arcsec^{-1}$.

These considerations made M87 a natural target for observations with the Near Infrared Camera and Multiobject Spectrometer (NICMOS) following its installation on HST in February 1997. As part of the NICMOS Guaranteed Time Observer program, we thus obtained images of the core region of the galaxy in the *JHK*-analog filters, as well as a medium-band filter centered at 2.37 μm and covering the stellar CO (2,0) bands (see, e.g., Oliva et al. 1999; Ivanov et al. 2000). This study is a companion to one we completed for the cores of two other nearby galaxies with evidence of nuclear black holes, M31 and M32 (Corbin, O'Neil & Rieke 2001; hereafter COR01), in which NICMOS images in the *JHK*-analog filters are also presented. As in COR01, we further compare our F160W-band image to archival Wide Field Planetary Camera 2 (WFPC2) images taken in the *V* and *I* analog filters, F555W and F814, to examine the optical-infrared color of the central region of the galaxy and assess its dust content. The results of Quillen, Bower & Stritzinger (2000) have shown that HST F555W – F160W images are a good diagnostic of dust in the cores of elliptical galaxies. This is particularly important for M87 as its proximity in principle allows the spatial resolution of the putative dusty torus surrounding the accreting black hole: unified models of quasars and radio galaxies predict that this torus may extend ~ 20 – 200 pc for a hole ~ $10^9$ $M_\odot$ (Urry & Padovani 1995). Strong mid-infrared emission from such a torus in M87 has not been detected (Perlman et al. 2001a). In the following section, we discuss our observations and reductions. In § 3 we present our results, and conclude with a discussion of them in § 4.

## 2. OBSERVATIONS AND REDUCTIONS

Our NICMOS observations of M87 as well as information on the archival WFPC2 *V* band (F555W) image are summarized in Table 1. The images were taken with NICMOS camera 2 in MULTIACCUM mode and in a four-step spiral dither pattern. The field size of NICMOS camera 2 is 19.2″, which covers the central region of M87 out to a distance of approximately 750 pc from the nucleus at the adopted distance to the galaxy. We performed what we believe to be an optimum reduction of the raw images through the use of a set of "super" flat-field and dark current frames, created from series of on-orbit frames taken close in time to these observations. We also performed a sky subtraction for each of the final broadband images using a mean sky value derived from a set of NICMOS camera 2 parallel fields (see Corbin et al. 2000), which should provide a good approximation to the sky level under the galaxy. The images were flux calibrated from observations of standard stars made on-orbit (Colina & Rieke 1997)[2] , and we estimate these calibrations to be accurate to within approximately 10%. No correction for Galactic extinction was made, as the maps of Schlegel, Finkbeiner & Davis (1998) indicate that the extinction values at the position of the galaxy in the near infrared are negligible. The orientation of all the NICMOS images (degrees by which the images are rotated east of north) is 78.10º. The particular WFPC2 F555W image selected from the archive was chosen on the basis of having comparable depth to the NICMOS images, and for being nearest in time to them.

We did not attempt to deconvolve the images from their point-spread functions (PSFs), as such deconvolution has proven problematic for NICMOS images, particularly without the use of a star specifically observed for determination of the PSF (see COR01). We did, however, use simple Gaussian convolution of the NICMOS F160M image to match its PSF to that of the WFPC2 F555W image as best as possible, and likewise convolve the NICMOS F110W and F160W images to match the wider PSF of the F222M image for the purpose of measuring the colors in those bands. In all other cases the images were not altered, in order to preserve their resolution.

Standard stars were observed in the NICMOS F110W and F160W filters on the same date that the M87 images in these filters were obtained. We experimented with scaling and subtracting these standard stars from the M87 nucleus with the hope that the results would reveal more structure in the nuclear region. However, for both the F110W and F160W images the results of these subtractions showed strong residual structure at the scale of the image point-spread functions, which we attribute to a combination of effects including differences between the orientation angles at which the stars and M87 were observed (even after rotation and alignment), and differences between the positions of the stars and the M87 nucleus on the detector. We are thus unable to obtain reliable information from these images about the nuclear region of the galaxy at scales less than ~ 0.5″.

---

[2] see also http://www.stsci.edu/instruments/nicmos

# 3. RESULTS

## 3.1 Images

A color composite of the NICMOS F110W, F160W and F222M images is shown in Figure 1. The faint red ripples are a consequence of the wider PSF of the F222M image. The red spot at the upper left of the image is the NICMOS camera 2 coronagraphic hole, which is emissive in the F222M filter. In addition to the basic symmetry of the galaxy, the notable feature of the image is the blueness of the globular clusters relative to the underlying galaxy. We present more details on this result below. There is also no evidence of any localized or circumnuclear star formation.

Figure 2 presents the F237M image. The coronagraphic hole is again evident. This image lacks the depth of the broadband images, as evidenced by the residual flat-field pattern at its edges. The important feature of this image however is the lack of a strong gradient in intensity away from the nucleus, indicating that there is no strong gradient in the strength of the stellar CO (2,0) bands. As a further check of this result, we took the ratio of this image to the F222M image, which covers the adjacent continuum, and also found no evidence of a radial gradient, or of any regions of enhanced CO (2,0) emission. These results agree with the lack of broadband color gradients in the region of the galaxy imaged, which we present in § 3.2.

In Figure 3 we present contour plots of the F110W, F160W and F222M images. The nearly perfectly circular nature of the isophotes is striking. We began the isophotes approximately 0.5″ (50 pc) from the nuclear peak, to avoid the Airy rings of the nuclear PSF. Nonetheless, even at such small distances there is no evidence of a departure from an almost perfectly symmetric distribution of the galaxy light (the small kinks in the isophotes in the vertical direction and at the midpoint of the plots are an artifact of the boundaries between the detector quadrants, which cannot be removed in the reduction process).

As a further test of the remarkable circularity evident in the contour plots, we measured the ellipticity of the isophotes using the IRAF.STSDAS task "ellipse," in which the ellipticity of each isophote is defined as [1 – (minor axis / major axis)]. This was done after first removing the jet knots, globular clusters and coronagraphic hole from the images with quadratic interpolation between the surrounding pixels. The results are shown in Figure 4, which compares the ellipticity measurements in the F110W, F160W and F222M filters versus distance along the semi-major axis. This plot shows that even with the imperfect removal of the jet knots and coronagraphic hole, the deviation of the isophotes from perfect circularity is at most ~ 10% and is typically ~ 1% - 2% beyond approximately 3″ from the nucleus.

Figure 5 shows the difference image formed by subtracting the NICMOS F160W image from the WFPC2 F555W image. This image was produced after first re-sampling the F160W image to the same pixel scale as the F555W image, then rotating and aligning the images at the position of the nucleus. As noted in § 2, we also iteratively convolved the F160W image to better match the resolution of the F555W image, as determined by the PSF of the nucleus. Obtaining quantitative information from this difference image is problematic, as the PSF matching process is imperfect and can introduce small gradients into the final image (see COR01). Somewhat surprisingly, the image does not clearly show the faint dust features seen in the emission-line and continuum images of Sparks et al. (1993), Ford et al. (1994) and Sparks et al. (2000), although other color differences, e.g. in the jet, are evident. Given that the dust in this region is not strongly concentrated, this may be an artifact of the imperfect PSF matching, i.e. diffuse features have been smeared out. Contamination of the optical continuum emission by the [O III] λλ4959, 5007 lines is also a concern, as the spectroscopy of Harms et al. (1994) indicates that they may contribute up to ~ 70% of the emission in this band near the M87 nucleus. We therefore also compared the F160W image with an archival WFPC2 F814W image, which although closer in wavelength to the F160W image is free of strong line emission. The result was very similar to the F555W – F160W image, in that no strong dust features are seen, although patches of dust show up faintly in the F814W and F555W images alone. Emission in [O III] λλ4959, 5007 is thus not likely to dominate the F555W – F160W image, and we conclude that the M87 core does not contain concentrated dust at the level of some of the early-type galaxies in the sample of Quillen et al. (2000), or many low-redshift radio galaxies (e.g. Martel et al. 1999, 2000; Perlman et al. 2001c).

There is no clear evidence in Figure 5 (or in the F814W – F160W image) of a compact torus of dust surrounding the nucleus, which if physically thick would show up as an ellipse around the nucleus that is lighter than the surrounding stars. However, the Ford et al. (1994) images show the [O III] λ5007 emission to be strong within 1″ of the nucleus, so the detection of such a torus is again subject to the amount of contamination in the F555W image by this line (as well as the residual differences between the image PSFs). Assuming however that the detection of a torus were possible with these data, then Figure 5 indicates its outer radius to be < 50 pc. The F555W – F160W colors of the jet knots are qualitatively consistent with synchrotron-dominated emission, as found from HST WFPC2 and NICMOS images covering the nucleus and further along the jet axis by Perlman et al. (2001b).

## 3.2 Color and Brightness Profiles

In Figure 6 we present the colors formed from the F110W, F160W and F222M images, azimuthally averaged around the nucleus, and measured on a Vega magnitude scale. These colors were begun at a distance large enough to avoid the nuclear peak and the region affected by the differences in the filter PSFs, and traced out to a distance of approximately 5″ (400 pc) from the nucleus, beyond which the associated errors became very large. These colors are clearly constant to within the associated errors. Using the stellar spectral library of Bruzual as implemented in the IRAF.STSDAS package "synphot" (White et al. 1998), we find that these colors are best matched by M4 III - M5 III stars, and there is thus no strong evidence of recent star formation. The bright globular cluster located approximately 6″ south of the nucleus has colors bluer than these values by ~ 0.2 - 0.3 magnitudes, depending on the amount of contamination from the underlying galaxy, and best corresponds to a population dominated by late G / early K main sequence stars. This is consistent with the study of the globular clusters in the core of the galaxy by Kundu et al. (1999), which reveals a population of relatively blue clusters in the region.

The azimuthally-averaged galaxy brightness profile of the F160W image (from which the jet emission and globular clusters have been removed) is shown in Figure 7. Motivated by the results of Lauer et al.

(1992), who find that an $r^{1/4}$ - law fits the deconvolved *V* and *I*-band HST Planetary Camera images of M87 within 3″ of the nuclear peak, we applied a de Vaucouleurs $r^{1/4}$ - law of the form

$$I(r) = I_0 \exp\{-7.688[(r/r_0)^{1/4} - 1]\}$$

to this brightness profile, using the IRAF.STSDAS task “nfit1d,” and found that a good fit can be obtained. Similar profiles and fits were obtained for the F110W and F222M images. This result is consistent with that of Lauer et al. (1992), and provides further evidence of a cusp in the stellar light associated with the nuclear black hole, insofar as there is no evidence of a flattening of the brightness profile down to scales ~ 50 pc.

## 4. DISCUSSION

### 4.1 The Dynamical State of the M87 Core

The strong circular symmetry of the stellar isophotes of M87 (Figs. 3 and 4) and the approximate fit of a simple functional form to its brightness profile (Fig. 6) is consistent with the models of Merritt & Quinlan (1998) that supermassive black holes can, depending on the mass of their host galaxy, produce a nearly spherical shape in the stellar distribution on the order of a galaxy crossing time. This symmetry is upon first consideration inconsistent with the evidence of the kinematic irregularities found by Pierce & Berrington (2000) for the stars in this region. However, these kinematic irregularities are ~ 10%, in terms of their deviation from a pure Gaussian model of the velocity cross-correlation function, and thus may not have a strong photometric signature. The Pierce & Berrington (2000) results may also be affected by dust extinction within the galaxy. Such spherical distributions can also retain small anisotropies in the stellar kinematics associated with the hole, as found for M87 by Merritt & Oh (1997). However, additional evidence that M87 formed from the merger of two or more galaxies is provided by the bimodal color distribution of its innermost globular clusters, suggesting different formation histories possibly related to merging (Kundu et al. 1999). Kobayshi & Arimoto (1999) also find very small metallicity gradients within M87 and other elliptical galaxies, which is inconsistent with monolithic collapse models of their formation. Our data are consistent with these results, given that we find no evidence of a stellar population gradient within the core region, and we confirm that the globular clusters in this region have different colors than the underlying galaxy. Similarly, the lack of evidence of "stellar wakes" produced by separate black holes in a coalescence model for the nuclear black hole also argues against recent formation for the latter, although again the photometric strength of such features is uncertain and the time scales, particularly for the accretion of low-mass galaxies by ellipticals, are short (see, e.g., Hernquist & Quinn 1988).

### 4.2 Constraints on the Nuclear Torus

The lack of clear evidence of a dust torus from the F555W – F160W and F814W – F160W images is consistent with the lack of strong thermal mid-infrared emission from the nucleus reported by Perlman et al. (2001a). This is in contrast to evidence of such tori in other nearby radio galaxies including Centaurus A (Rydbeck et al. 1993; Mirabel et al. 1999) and Cygnus A (Conway 1999; Radomski et al. 2002), both of which have ratios of nuclear mid-infrared to radio emission ~ $10^3$ times higher than that observed in M87. However, the size limit we infer for the M87 torus is consistent to within an order of magnitude with the theoretical predictions for its size (see § 1), so a torus may still be present but is small relative to the Cen A and Cyg A tori. A more speculative interpretation would be that this torus, as well as the broad-line region that it formerly shielded, is being accreted by the black hole as part of the last phase of its activity.

### 4.3 Comparison to Other Radio Galaxies and the Evolutionary State of the Nucleus

HST observations of other radio galaxies including 3C sources (e.g. Martel et al. 1999; 2000; Sparks et al. 2000; Perlman et al. 2001c) have revealed them to typically contain large-scale (~ $10^2$ – $10^3$ pc) dust disks oriented roughly perpendicular to their jet axes. Our results confirm that M87 differs from these objects in that no evidence of such a disk is seen in either the NICMOS images or the associated colors. M87 stands in particular contrast to its local counterparts Cen A and Cyg A, not only in the mid-infrared properties of its nucleus but in terms of its large-scale dust content. Taken together, these results suggest that M87 is older than other local radio galaxies, or at least underwent any major mergers related to its nuclear activity longer ago. In addition to its lack of strong mid-infrared emission, two facts also suggest that the M87 nucleus is in the last stages of its activity. First, the estimated ~ $10^9 M_{\odot}$ mass of the nuclear black hole is comparable to those estimated for luminous quasars (e.g. Laor 2000), yet the M87 nuclear luminosity is lower by several orders of magnitude, indicating a comparable difference in its accretion rate. Second, the lack of recent star formation in the nuclear region evidenced by the galaxy's colors (§ 3.2) will also deprive the central black hole of mass shed by early-type stars, which has been proposed as an important source of fuel to AGNs (Norman & Scoville 1988). Without a major "refueling" event such as the cannibalism of a smaller galaxy in the Virgo cluster, the M87 nucleus thus appears to be on its way to quiescence.

We thank the referee, Eric Perlman, for his helpful review of this paper. We also thank Valentin Ivanov, Matthias Steinmetz, Betty Stobie and Glenn Schneider for their help during the course of this project. This work was supported in part by NASA grants NAG 5-3042 and GO-08333.01-97A, the latter from the Space Telescope Science Institute, which is operated by AURA, Inc., under NASA contract NAS5-26555, and in part by Computer Sciences Corporation. This work has also made use of the NASA Extragalactic Database (NED), which is supported by the Jet Propulsion Laboratory under contract with NASA.

REFERENCES

Cohen, J.C. 2000, AJ, 119, 162

Colina, L. & Rieke, M. 1997, in The 1997 HST Calibration Workshop, eds. S. Casertano, R. Jedrzejewski, C.D. Keyes & M. Stevens (Baltimore: STScI), 182

Conway, J.E. 1999, in Highly Redshifted Radio Lines, ASP Conference Series 156, eds. C.L. Carilli, S.J.E. Radford, K.M. Menten, & G.I. Langston (San Francisco: ASP), 259

Corbin, M.R., O'Neil, E.J., Thompson, R.I., Rieke, M.J. & Schneider G. 2000, AJ,120, 1209

Corbin, M.R., O'Neil, E.J. & Rieke, M.J. 2001, AJ, 121, 2549 (COR01)

Faber, S.M. et al. 1997, AJ, 114, 1771

Ford, H.C. et al. 1994, ApJ, 435, L27

Harms, R.J. et al. 1994, ApJ, 435, L35

Hernquist, L., & Quinn, P.J. 1988, ApJ, 331, 682

Ivanov, V., Rieke, G.H., Groppi, C.E., Alonso-Herrero, A., Rieke, M.J. & Engelbracht, C.W. 2000, ApJ, 545, 190

Kobayashi, C. & Arimoto, N. 1999, ApJ, 527, 573

Kundu, A., Whitmore, B.C., Sparks, W.B., Macchetto, F.D., Zepf, S.E. & Ashman, K.E. 1999, ApJ, 513, 733

Laor, A. 2000, ApJ, 543, L111

Lauer, T.R. et al. 1992, AJ, 103, 703

Macchetto, F., Marconi, A., D.J., Axon, Capetti, A., Sparks, W. & Crane, P. 1997, ApJ, 489, 579

Marconi, A., Axon, D.J., Macchetto, F.D., Capetti, A., Sparks, W.B. & Crane, P. 1997, MNRAS, 289, 21

Martel, A. et al. 1999, ApJS, 122, 81

Martel, A.R., Turner, N.J., Sparks, W.B. & Baum, S.A. 2000, ApJS, 130, 267

Merritt, D. & Oh, S.P. 1997, ApJ, 113, 1279

Merritt, D. & Quinlan, G.D. 1998, ApJ, 498, 625

Mirabel, I.F. et al. 1999, A&A, 341, 667

Norman, C. & Scoville, N. 1988, ApJ, 332, 124

Oliva, E., Origlia, L., Maolino, R. & Moorword, A.F.M. 1999, A&A, 350, 9

Perlman, E.S., Sparks, W.B., Radomski, J., Packham, C., Fisher, R.S., Piña, R. & Biretta, J.A. 2001a, ApJ, 561, L51

Perlman, E.S., Biretta, J.A., Sparks, W.B., Macchetto, F.D. & Leahy, J.P. 2001b, ApJ, 551, 206

Perlman, E.S., Stocke, J.T., Conway, J. & Reynolds, C. 2001c, AJ, 122, 536, 548

Pierce, M., Welch, D.L., McClure, R.D., van den Bergh, S., Racine, R. & Stetson, P.B. 1994, Nature, 371, 385

Pierce, M.J. & Berrington, R.C. 2000, ApJ, 531, L99

Quillen, A., Bower, G.A. & Stritzinger, M. 2000, ApJS, 128, 85

Radomski, J.T., Piña, R.K., Packham, C., Telesco, C.M. & Tadhunter, C.N. 2002, ApJ, 566, 675

Rydbeck, G., Wiklind, T., Cameron, M., Wild, W., Eckart, A., Genzel, R., & Rothermel, H. 1993, A&A, 270, L13

Schlegel, D.J., Finkbeiner, D.P. & Davis, M. 1998, ApJ, 500, 525

Sparks, W.B., Ford, H.C. & Kinney, A.L. 1993, ApJ 413, 531

Sparks, W.B., Baum, S.A., Biretta, J., Macchetto, F.D. & Martel, A.R. 2000, ApJ, 542, 667

Urry, C.M. & Padovani, P. 1995, PASP, 107, 803

White, R. et al. 1998, Synphot User's Guide (Baltimore: STScI)

Whitmore, B.C., Sparks, W.B., Lucas, R.A., Macchetto, F.D. & Biretta, J.A. 1995, ApJ, 454, L73

Young, P., Westphal, J., Kristian, J. Wilson, C. & Landauer, F. 1978, ApJ, 221, 721

TABLE 1

SUMMARY OF OBSERVATIONAL DATA

| Instrument | Filter | Central Wavelength (μm) | Integration Time (s) | UT Date (m/d/y) |
|---|---|---|---|---|
| NICMOS | F110W | 1.102 | 96 | 11/20/97 |
| “ | F160W | 1.593 | 32 | “ |
| “ | F222M | 2.216 | 128 | “ |
| “ | F237M | 2.369 | 192 | “ |
| WFPC2 | F555W | 0.525 | 230 | 5/27/95 |

FIGURE CAPTIONS

FIGURE 1. – Color composite of the F110W, F160W and F222M filter images of the M87 core, where these images have been loaded into the blue, green and red color channels, respectively. The red spot in the upper left of the image is the camera 2 coronagraphic hole, which is thermally emissive in the F222M filter.

FIGURE 2. -- Gray-scale plot of the F237M filter image. The lighter regions at the edges of the image are the result of imperfect flat fielding. The dark spot at the upper left is the coronagraphic hole.

FIGURE 3. – Contour plots of the F110W, F160W and F222M filter images. Image orientation is the same as Figures 1 and 2. Contours were begun well below the nuclear peak, and are on a logarithmic scale with intervals of 0.05 dex. The log of the flux values of the innermost contours are (in ergs $s^{-1}$ $cm^{-2}$ $Å^{-1}$) -17.55 (F110W), -17.65 (F160W) and -18.40 (F222M). The coronagraphic hole is again visible in the F222M image.

FIGURE 4. – Comparison of the ellipticity parameter and distance along the semi-major axis for the F110W, F160W and F222M images, after removal of the jet knots, globular clusters and coronagraphic hole by interpolation. The excursions in these values near 0.7″ and 2″ are due to the imperfect removal of the jet knots, and in the case of the F222M image the imperfect removal of the coronagraphic hole produces the excursions at 11″ - 12″.

FIGURE 5. – F555W - F160W difference image. Darker regions are bluer, lighter regions are redder.

FIGURE 6. – Azimuthally-averaged NICMOS color profiles of the inner 5″ of M87. Values are on a Vega magnitude scale and represent averages within concentric circular annuli of 0.2″ centered at the nucleus. Error bars represent the $1\sigma$ deviation of individual pixels within each annulus.

FIGURE 7. -- Azimuthally-averaged brightness profile of the F160W image (open squares). The solid line represents the best fit of a de Vaucouleurs $r^{1/4}$ - law to the profile.